\documentclass[acmsmall,screen,review,table,nonacm]{acmart}
\renewcommand\footnotetextcopyrightpermission[1]{}
\makeatletter
\renewcommand\received[2][]{}
\makeatother

\usepackage{graphicx}     
\usepackage{subcaption}   
\usepackage{mwe} 
\usepackage{enumitem}
\usepackage{multirow}
\usepackage{wrapfig}

\newcommand{\todo}[1]{{\textcolor{black}{#1}}}
\newcommand{\chong}[1]{{\textcolor{orange}{}}}

\AtBeginDocument{%
  }

\setcopyright{acmlicensed}
\copyrightyear{2018}
\acmYear{2018}
\acmDOI{XXXXXXX.XXXXXXX}
\acmConference[Conference acronym 'XX]{Make sure to enter the correct
  conference title from your rights confirmation email}{June 03--05,
  2018}{Woodstock, NY}
\acmISBN{978-1-4503-XXXX-X/2018/06}

\begin{document}

\title{Exploring and Unleashing the Power of Large Language Models in CI/CD Configuration Translation }
\author{Jiajun Wu}
\authornote{Jiajun Wu and Chong Wang have equal contributions (co-first authors).}
\email{jjwu123@stu.suda.edu.cn}
\affiliation{%
  \institution{Soochow University}
  \country{China}
}
\author{Chong Wang}
\authornotemark[1]
\email{chong.wang@ntu.edu.sg}
\affiliation{%
  \institution{Nanyang Technological University}
  \country{Singapore}
}

\author{Chen Zhang}
\authornote{Chen Zhang is the corresponding author.}
\affiliation{%
  \institution{Soochow University}
  \country{China}}
\email{chenzhang22@suda.edu.cn}

\author{Wunan Guo}
\affiliation{%
  \institution{University of Shanghai for Science and Technology}
  \country{China}
}
\email{wnguo@usst.edu.cn}

\author{Jianfeng Qu}
\affiliation{%
 \institution{Soochow University}
 \country{China}
}
\email{jfqu@suda.edu.cn}

\author{Yewen Tian}
\affiliation{%
 \institution{Soochow University}
 \country{China}
}
\email{ywtian08175452@stu.suda.edu.cn}

\author{Yang Liu}
\affiliation{%
  \institution{Nanyang Technological University}
  \country{Singapore}
}
\email{yangliu@ntu.edu.sg}
\renewcommand{\shortauthors}{Wu et al.}

\begin{abstract}

Continuous Integration (CI) is a cornerstone of modern collaborative software development, and numerous CI platforms are available. Differences in maintenance overhead, reliability, and integration depth with code-hosting platforms make migration between CI platforms a common practice. A central step in migration is translating CI configurations, which is challenging due to the intrinsic complexity of CI configurations and the need to understand semantic differences and relationships across CI platforms.

With the advent of large language models (LLMs), recent advances in software engineering highlight their potential for CI configuration translation. In this paper, we present a study on LLM-based CI configuration translation, focusing on the migration from Travis CI to GitHub Actions. First, using 811 migration records, we quantify the effort involved and find that developers read an average of 38 lines of Travis configuration and write 58 lines of GitHub Actions configuration, with nearly half of the migrations requiring multiple commits. We further analyze translations produced by each of the four LLMs and identify 1,121 issues grouped into four categories: logic inconsistencies (38\%), platform discrepancies (32\%), environment errors (25\%),  and syntax errors (5\%). Finally, we evaluate three enhancement strategies and show that combining guideline-based prompting with iterative refinement achieves the best performance, reaching a Build Success Rate of 75.5\%—nearly a threefold improvement over GPT-4o with a basic prompt.

\end{abstract}


\settopmatter{printacmref=false}
\renewcommand\footnotetextcopyrightpermission[1]{}
\pagestyle{plain}

\received{20 February 2007}
\received[revised]{12 March 2009}
\received[accepted]{5 June 2009}

\maketitle

\section{Introduction}\label{sec:intro}

Continuous Integration (CI) is a cornerstone of modern collaborative software development~\cite{golzadeh2022rise}. It automatically builds code changes (e.g., compilation, static analysis, and testing) to identify potential bugs before integration into the central repository. CI has been widely adopted for its ability to detect bugs early and accelerate development cycles, thereby improving software reliability, quality, developer productivity, and process transparency~\cite{hilton2017trade, rostami2023usage}. 
With the growing importance of CI, its ecosystem has expanded substantially. According to a recent survey~\cite{rostami2023usage}, numerous CI platforms such as GitHub Actions~\cite{github-action-com} and Travis CI~\cite{travis-ci-com} are in use.

Central to CI workflows is the configuration, which defines the build environment, required dependencies, and build commands. Most CI configurations are expressed in YAML syntax~\cite{rostami2023usage}. For instance, Travis CI uses a \textit{.travis.yml} file located at the repository root, whereas GitHub Actions stores configuration files under \textit{.github/workflows/} with \textit{.yml} or \textit{.yaml} extensions. Due to the inherent complexity of CI configurations~\cite{hilton2017trade}, working with YAML format presents challenges such as confusing and error-prone syntax and semantics, difficulty in testing and debugging, limited auto-completion support, and inadequate documentation~\cite{saroar2023developers, zhang2024developers}. Misconfigured CI workflows can adversely affect correctness, maintainability, performance, and security~\cite{gallaba2018use, vassallo2020configuration, zhang2022buildsonic, khatami2024catching}.

Although CI platforms generally offer similar functionalities, migrations between them are a common practice~\cite{golzadeh2022rise, rostami2023usage} due to differences in important aspects such as maintenance overhead, reliability, and integration depth with code hosting platforms. Such migrations require transforming configurations from one CI platform to another while preserving the original workflows' behavior, a process we term \textbf{CI configuration translation}. Performing this translation manually is challenging, not only due to the intrinsic difficulties of CI configuration, but also because it demands a deep understanding of the semantic differences and relationships between the two CI platforms' configuration rules~\cite{rostami2023usage}. Therefore, automated solutions for translating CI configurations across platforms are essential. To address this challenge, GitHub Actions provides an official tool, \textit{Importer}, to facilitate configuration translation from other CI platforms~\cite{importer}. However, it uses a rule-based approach relying on predefined manual mappings~\cite{rzig2024example}. This limits its adaptability, as it requires continuous updates whenever GitHub Actions or the source CI platform evolves, and is difficult to generalize to other CI platforms. Moreover, it often struggles to generate idiomatic configurations, which are harder to maintain. Rzig et al.~\cite{rzig2024example} introduced an example-based mining tool, \textit{CIMig}, which automatically infers translation rules from existing migration examples. However, it requires high-quality migration examples for automatic mining, and its accuracy remains unsatisfactory~\cite{nazmul2025cigrate}.

With the advent of large language models (LLMs), recent research has successfully applied them to diverse software engineering tasks, including code generation~\cite{du2024evaluating}, vulnerability detection~\cite{zhou2024large}, and code translation~\cite{pan2024lost}, demonstrating impressive results. These advances suggest that LLMs can potentially overcome the limitations of existing CI migration tools. It is thus important to investigate the effectiveness of LLMs for the automatic CI configuration translation.
In this work, we conduct a study to bridge key gaps in understanding LLM-based cross-platform CI configuration translation.
Specifically, we focus on migration from Travis CI to GitHub Actions, a common migration pattern, as Travis CI was once the dominant CI platform for open-source projects but has largely been supplanted by GitHub Actions~\cite{golzadeh2022rise}. First, we quantify the effort required to achieve successful migration, thereby motivating the need for automated translation support.
Second, we evaluate the fundamental ability of LLMs in CI configuration translation and develop a taxonomy of translation issues.
Finally, we investigate the feasibility of straightforward enhancement strategies, including one-shot prompting, guideline-based prompting, and iterative refinement using CI build feedback. Guided by these objectives, we address the following three research questions.

\noindent \textbf{RQ1: How much effort is required to achieve a successful translation from Travis CI to GitHub Actions?}

\noindent \underline{Motivation.} Migrating CI is a non-trivial task, but there is still a lack of quantitative data to characterize its effort. Understanding this highlights the need for automated CI configuration translation techniques.

\noindent \underline{Results.} On average, developers read 38 lines of configuration and write 58 lines to achieve a successful translation. Approximately 47.7\% of translations require more than one commit. 
Modification commits typically involve 12 lines added and 7 lines deleted.

\noindent \textbf{RQ2: What types of translation issues occur in LLM-generated configurations?}

\noindent \underline{Motivation.} While prior research~\cite{pan2024lost, yang2024exploring, du2024evaluating} has examined common failures in cross-language code translation, CI configuration translation presents distinct characteristics. Both source and target configurations use the same flexible YAML syntax (a simple key--value format), but they differ significantly in configuration semantics and rules. Characterizing these issues is essential to understanding the limitations of LLM-generated configurations and to informing the design of more effective translation techniques.

\noindent \underline{Results.} Our analysis identified 1,121 issues in the translated configurations, grouped into four categories: syntax errors, platform discrepancies, environment errors, and logic inconsistencies. Logic inconsistencies were the most frequent (38\%), followed by platform discrepancies (32\%), environment errors (25\%), and syntax errors (5\%).

\noindent \textbf{RQ3: To what extent can enhancement strategies improve the performance of LLM-based CI configuration translation?}

\noindent \underline{Motivation.} Based on the results of RQ2, the fundamental capability of LLMs in CI configuration translation remains limited. We therefore propose three enhancement strategies to assess whether they can improve translation performance.

\noindent \underline{Results.} The combined guideline-based prompting and iterative refinement strategy achieved the best performance, with a BSR of 75.5\% (173 successful translations), representing nearly a threefold improvement over the basic LLM baseline of 25.8\%.

In summary, this work makes the following contributions.

\begin{itemize}[leftmargin=*]
\item We construct a dataset of real-world CI migration records from Travis CI to GitHub Actions.

\item We evaluate the performance of four representative LLMs on CI configuration translation and develop a taxonomy of translation issues.

\item We propose and evaluate three enhancement strategies, demonstrating that guideline-based prompting combined with iterative refinement significantly improves the Build Success Rate.

\end{itemize}

\section{Background}\label{sec:background}

\textbf{CI/CD.} In modern software development, Continuous Integration (CI) and Continuous Delivery/Deployment (CD) are commonly discussed together under the umbrella term CI/CD. The key difference lies in the deployment phase, where CD extends CI by automating the software release process. Most CI platforms support both CI and CD. For simplicity, we use the term CI in this paper to broadly refer to both CI and CD. A typical CI \textit{build} (or \textit{workflow run}) is triggered when developers push code changes to a repository. The CI server automatically detects them and triggers one or more \textit{jobs}. Each job can be configured to run under different environments, such as varying operating systems or dependency versions. A job typically consists of sequential steps such as cloning the repository, checking out code, compiling, and running automated tests. Once all jobs are complete, feedback is provided to the development team, indicating whether the build succeeded or failed. This continuous feedback loop facilitates early defect detection and accelerates development.

With the growing importance of CI, a wide variety of platforms has emerged, ranging from self-hosted systems to managed services~\cite{buildkite-com}. Self-hosted platforms (e.g., Jenkins~\cite{jenkins-io}) require teams to install, configure, and maintain CI on their own servers. Managed platforms (e.g., GitHub Actions and Travis CI) are cloud-based services that abstract away most infrastructure details. Migrating between CI platforms typically involves multiple steps, such as provisioning servers, setting up environments, and translating configuration files. In this study, we focus specifically on automatic solutions for configuration translation to reduce the burden of CI migration.


\textbf{CI Configuration.} YAML is a human-readable data serialization language widely used for tasks such as configuration files, internet messaging, object persistence, data auditing, and visualization~\cite{yaml-org}. A YAML document is fundamentally composed of key--value mappings, where values can be scalars, sequences, or nested mappings. The simplest value type is a scalar, which represents zero or more Unicode characters, such as strings, numbers, or booleans. A value may also be a sequence, i.e., an ordered list of items denoted by a hyphen followed by a space. More complex structures can be expressed as mappings, where each value is defined as a set of key–value pairs. YAML relies on indentation to represent hierarchy and nesting, and it supports comments that begin with the hash symbol (\texttt{\#}).

\begin{figure}[!t]
    \centering
    \includegraphics[width=1\textwidth]{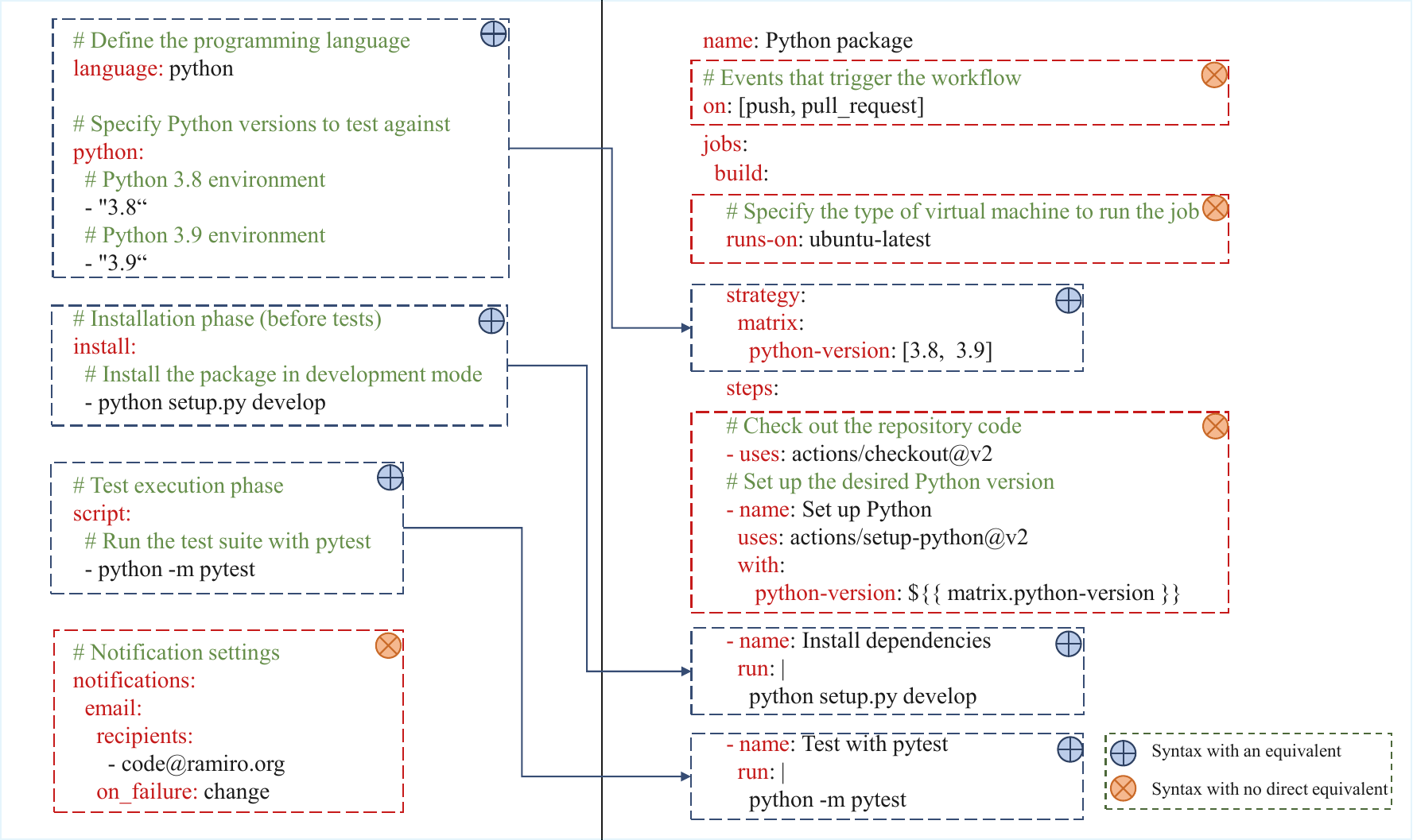}
    \vspace{-8pt}
    \caption{Example Configurations of Travis CI and GitHub Actions}
    \label{fig:configuration-excerpts}
\end{figure}

Most CI platforms adopt YAML as their configuration language due to its concise syntax and readability. Despite YAML’s common adoption, different CI platforms define their own semantics, so configurations are not directly interchangeable. Figure~\ref{fig:configuration-excerpts} shows example configurations in Travis CI (left) and GitHub Actions (right) that achieve the same behavior. In the figure, the blue $\oplus$ symbol highlights configurations with direct equivalents across the two platforms, whereas the orange $\otimes$ symbol indicates configurations without direct counterparts.
The Travis CI configuration specifies Python as the project language (\texttt{language: python}) and defines two jobs targeting versions 3.8 and 3.9. It defines the \texttt{install} phase as running \texttt{python setup.py develop} and the \texttt{script} phase as executing \texttt{python -m pytest}. It also includes a notification setting to send messages to a designated email account.

The GitHub Actions configuration defines a job matrix for Python 3.8 and 3.9, with steps that install the package (\texttt{python setup.py develop}) and run tests (\texttt{python -m pytest}). However, several aspects implicit in Travis CI must be explicitly specified in GitHub Actions. For example, developers must explicitly define trigger events (\texttt{on: [push, pull\_requests]}), use \texttt{actions/checkout} to check out the repository code, and use \texttt{actions/setup-python} to configure the Python runtime. Both \texttt{actions/checkout} and \texttt{actions/setup-python} are reusable \textit{actions} invoked with the \texttt{uses} keyword. These steps are implicit in Travis CI, either managed through the web interface or handled by default without explicit configuration. In addition, Travis CI provides support for notifications, enabling developers to configure email alerts directly in the configuration file. By contrast, GitHub Actions integrates notification features into the platform itself, so no explicit configuration is required in the workflow file.

\textbf{LLMs.} Language models are designed to simulate the human brain's ability to perform various tasks, such as speech recognition, machine translation, and information retrieval~\cite{jelinek1998statistical, manning1999foundations, ceri2013introduction, blank2023large}. The rise of LLMs has been made possible by advances in computational power, breakthroughs in machine learning techniques, and the availability of large-scale training datasets~\cite{hou2024large}. Compared with traditional language models, LLMs are not only significantly larger in model size, but also demonstrate stronger capabilities in language understanding and generation. Software engineering is a discipline concerned with the development, implementation, and maintenance of software systems~\cite{ieee1983ieee}. Because many software engineering tasks can be reframed as data, code, or text analysis tasks, LLMs have found wide application in this domain~\cite{wang2022machine}. The field of LLMs evolves at a rapid pace, with new models and techniques emerging continuously~\cite{minaee2024large}. Existing LLMs vary in characteristics such as openness (open-source vs. proprietary) and specialization (general-purpose vs. code-oriented). General LLMs (e.g., GPT-4o~\cite{openai-com}) are designed to handle a broad range of natural language tasks, while Code LLMs (e.g., DeepSeek-Coder~\cite{deepseek-com}) specialize in code-related tasks.

To enhance the performance of LLMs, various prompting strategies have emerged, which are often collectively referred to as \textit{prompting engineering}~\cite{wang2024advanced}. These techniques aim to guide the model toward more accurate and context-aware outputs by framing the task with tailored instructions or examples. Prompting strategies have proven effective across different software engineering tasks~\cite{gao2023makes}. Beyond single-turn prompts, developers also explore iterative refinement approaches and the integration of additional contextual information to refine the LLMs' output~\cite{yang2024exploring, zhong2024debug, pan2024lost, du2024evaluating}. These findings motivate us to explore whether similar strategies can enhance the performance of CI configuration translation.
\section{Methodology}\label{sec:method}

In this section, we present the methodology of our study. We first describe the data preparation process and the metrics used to assess translation effort, and then detail the experimental setup for LLM-based CI configuration translation.

\subsection{Data Preparation}
\textbf{Project Filtering.} To create a dataset of migration instances from Travis CI to GitHub Actions, we first used the SEART GitHub Search Engine (seart-ghs)~\cite{dabic2021sampling} to obtain a list of candidate open-source projects from GitHub. We selected seart-ghs because it is continuously updated and offers 25 filtering attributes (e.g., main language, creation time, number of stars), which simplifies the process of sampling relevant projects. As of \todo{ February 17, 2025}, seart-ghs indexed 1,738,020 projects. We focused on Python projects, as Python is one of the most popular programming languages~\cite{tiobe-com}. This yielded \todo{321,263} projects. 
Considering that GitHub Actions was released in November 2019, we retained only projects with commits after this date, resulting in \todo{263,394} projects. We also excluded forked projects, as their data may duplicate that of upstream projects, resulting in 252,628 projects. 

For the \todo{252,628} selected projects, \todo{we used the GitHub API to examine their repositories and verify the existence of a non-empty .github/workflows directory, which is a prerequisite for running GitHub Actions workflows. To further ensure that GitHub Actions were correctly configured and actively used, we also queried the GitHub API to confirm the presence of at least one successful workflow run}. After this validation step, we got 12,502 projects.

\textbf{Translation Record Extraction.} Among the 12,502 projects actively using GitHub Actions, our goal was to identify commits that migrated from Travis CI to GitHub Actions. To this end, we examined the commit history of each project and filtered commits whose messages matched regular expressions such as "migrate.*travis", "move.*travis", "replace.*travis", or "switch.*travis". This filtering step yielded \todo{2,601} candidate commits. Although these regular expressions are not exhaustive and developers may sometimes conduct the migration without explicitly mentioning it in the commit messages, this step substantially reduces the search space by filtering out irrelevant commits and ensuring that the remaining candidates were highly likely to be true migration records. Since GitHub Actions allows multiple workflow files, we restricted our analysis to one-to-one translations in which a Travis CI configuration file is replaced by a single GitHub Actions workflow file. After this validation step, we identified 811 migration commits. 

Because CI migration may not succeed in a single attempt, we further examined the subsequent change history of the 811 commits to determine their final stable versions and to extract the corresponding target GitHub Actions configurations.This process produced 811 translation records, each comprising one source Travis CI configuration file and its corresponding target GitHub Actions configuration file. Using these 811 translation records, we analyzed the required effort to answer RQ1.
To assess the quality of translated configurations, it was necessary to execute the workflows. However, we found that while many workflows were runnable at the time of their original migration, they could no longer be executed in the current environment due to factors such as \todo{the use of deprecated APIs, reliance on obsolete features, or incompatibilities introduced by the evolution of GitHub Actions.} To address these issues, we retained only workflows that could still be executed successfully, or could be made runnable with minimal modifications \todo{(e.g., updating action versions or replacing deprecated syntax)}. Through this process, we ultimately obtained 229 translation records. Based on these 229 translation records, we addressed RQ2 and RQ3.

\subsection{Translation Effort Measurement}\label{sec-effort}

To address RQ1, we examined the effort required for developers to migrate CI configurations from Travis CI to GitHub Actions. Based on the 811 translation records, we quantified this effort using three metrics: configuration size, migration attempts, and change size. 

\begin{itemize}[leftmargin=15pt]

\item \textbf{Configuration Size.} This metric measures the number of lines in the original Travis CI configuration or the corresponding GitHub Actions configuration. It is inspired by the widely used code complexity metric, \textit{Lines of Code} (\textit{LOC})~\cite{yu2010survey}. During translation, developers must read and understand the source configuration while constructing the target one. Intuitively, larger configuration files require greater cognitive and manual effort.

\item \textbf{Migration Attempts.} This metric captures whether the migration was completed in a single commit or required multiple commits before the GitHub Actions configuration stabilized. A single-commit migration suggests that developers translated the configuration correctly on the first attempt, whereas multiple commits imply additional debugging and refinement.

\item \textbf{Change Size.} For translations requiring multiple commits, this metric measures the number of configuration lines modified per commit. Conceptually similar to configuration size, it is assessed at the commit rather than the file level, thereby capturing the incremental effort invested during the migration process.

\end{itemize}

In summary, configuration size reflects the complexity of the final configuration, whereas migration attempts and change size capture the effort involved in the migration process.

\subsection{LLM-based Configuration Translation}

\textbf{Studied LLM Selection.} We experimented with four representative models: GPT-4o~\cite{openai-com}, GPT‑4o mini~\cite{openai-com}, Qwen-3~\cite{qwen-ai}, and DeepSeek-Coder~\cite{deepseek-com}. Among them, three are general LLMs: GPT-4o and GPT-4o mini, both proprietary models developed by OpenAI, and Qwen-3, an open-source model. Additionally, because configuration translation is inherently a software engineering task, we also evaluated a code LLM, DeepSeek-Coder. Table~\ref{tab:llm-overview} summarizes the selected LLMs, listing their name, model type, release time, parameter size, number of pretraining tokens (Training Base), and the input/output context window limits (In/Out).  

\begin{figure}[t]
    \centering
    \includegraphics[width=0.45\textwidth]{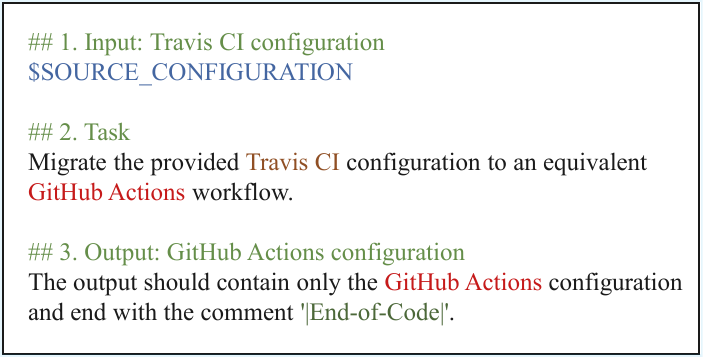}
    \caption{Basic Prompt Template}
    \label{fig:basic-prompt}
\end{figure}

\textbf{Prompt Template Design.} We designed a basic prompt template as shown in Figure~\ref{fig:basic-prompt} to evaluate the configuration translation capability of LLMs. The prompt consists of five components: (1) natural language instructions explicitly defining the translation task; (2) the source CI platform, Travis CI; (3) the target CI platform, GitHub Actions; (4) the configuration to be translated (\$SOURCE\_CONFIGURATION); (5) output control instructions that restrict the model to only generate the translated configuration, excluding extra explanations or descriptions. This instruction ensures that the output is clean and directly usable.


\textbf{LLM Feeding.} To ensure reproducibility, we set the \texttt{temperature} parameter to 0 for GPT-4o, GPT‑4o mini, and DeepSeek-Coder. In contrast, Qwen-3 was run with its default configuration, as setting the \texttt{temperature} to 0 leads to infinite loops and prevents valid response generation~\footnote{https://huggingface.co/Qwen/Qwen3-8B}. \todo{GPT-4o and GPT-4o mini were accessed via the official OpenAI API, whereas Qwen-3 was deployed on a server with an NVIDIA RTX 4090 (24 GB) GPU, and DeepSeek-Coder was executed on a server with an NVIDIA A10 (24 GB) GPU.} For all models, we instantiated the basic prompt template and replaced placeholders with concrete values for each translation record. Each translation record was processed exactly once.

\begin{table}[ht]
\centering
\caption{Overview of Studied LLMs}
\vspace{-10pt}
\small
\label{tab:llm-overview}
\begin{tabular}{cccccc}
\toprule
\textbf{Name} & \textbf{Type} & \textbf{Parameter Size} & \textbf{Release Time} & \textbf{Training Base} & \textbf{In/Out} \\
\midrule
GPT-4o & Proprietary &  \textasciitilde 200Billion & 2024-08-06 & - & 128k/16k \\
\hline
GPT-4o mini & Proprietary & - & 2024-07-18  & - & 128k/16k\\
\hline
Qwen-3 & Open-source & 8.2Billion & 2025-07-27  & -& -\\
\hline
DeepSeek-Coder & Open-source & 6.7Billion & 2025-02-27 & 2Trillion & -  \\
\bottomrule
\end{tabular}
\end{table}

\subsection{Translation Quality Measurement}

To evaluate the quality of translated configurations, we adopted one execution-based metric and two similarity-based metrics. 

\textbf{Execution-based Metric.} The execution-based metric assesses the quality of translated configurations based on their execution status.

\begin{itemize}[leftmargin=15pt]
\item \textbf{Build Success Rate (BSR).} This metric measures the ratio of translated configurations that can be successfully executed in the target CI environment. We define BSR as follows:
\begin{equation}
\text{BSR} = \frac{N_{success}}{N}
\label{eq:bsr}
\end{equation}

where $N$ denotes the total number of translated configurations, and $N_{success}$ represents the number that complete with a success status.
\end{itemize}

\textbf{Similarity-based Metrics.} These metrics evaluate translated configurations by comparing them with the ground-truth configurations using textual similarity, without requiring workflow execution. Following prior work~\cite{rzig2024example}, we employ Cosine Similarity~\cite{shalton1986introduction} and CrystalBLEU~\cite{eghbali2022crystalbleu}. \chong{Not sure if we need to introduce similarity metrics here, because they are not used in taxonomy construction?} 

\begin{itemize}[leftmargin=15pt]

\item \textbf{Cosine Similarity (CS).} This metric measures the similarity between the translated and ground-truth configurations in vector space. Each configuration is represented as a term-frequency vector, and the cosine of the angle between the two vectors is computed. The score ranges from 0 to 1, with higher values indicating stronger similarity.

\begin{equation}
\text{CS}(\vec{x}, \vec{y}) = \frac{\vec{x} \cdot \vec{y}}{\lVert \vec{x} \rVert  \lVert \vec{y} \rVert}
\end{equation}

where $\vec{x}$ and $\vec{y}$ denote the vector representations of the translated and ground-truth configurations, $\vec{x} \cdot \vec{y}$ is the dot product, and $\lVert \vec{x} \rVert$ and $\lVert \vec{y} \rVert$ are their Euclidean norms.

\item \textbf{CrystalBLEU (CB).} This metric is a variant of BLEU~\cite{papineni2002bleu}. It measures the $n$-gram overlap between translated and ground-truth configurations while excluding high-frequency boilerplate tokens. This adjustment prevents inflated scores caused by common keywords and yields a more reliable similarity measure.

\end{itemize}

\subsection{Manual Issue Annotation}
We construct a taxonomy of issues in translated CI configurations. For each LLM, we manually examined 229 generated configurations. The annotation was independently performed by two authors, one with seven years of CI experience and the other with three years. Following an open-coding procedure~\cite{Seaman1999QME}, the annotators analyzed the configurations to identify and label issues. When necessary, they consulted official documentation, developer forums, and workflow execution logs. In cases of disagreement, a third author with five years of CI experience was consulted to resolve conflicts and reach consensus. To evaluate the reliability of the coding process, we assessed inter-rater agreement. The resulting Cohen’s Kappa score was \todo{0.80}, indicating substantial agreement~\cite{Cohen1960A}.

\subsection{Enhancement Strategy Investigation}

\begin{figure}[htbp]
\centering
\begin{subfigure}[t]{0.33\textwidth}
    \includegraphics[width=\textwidth]{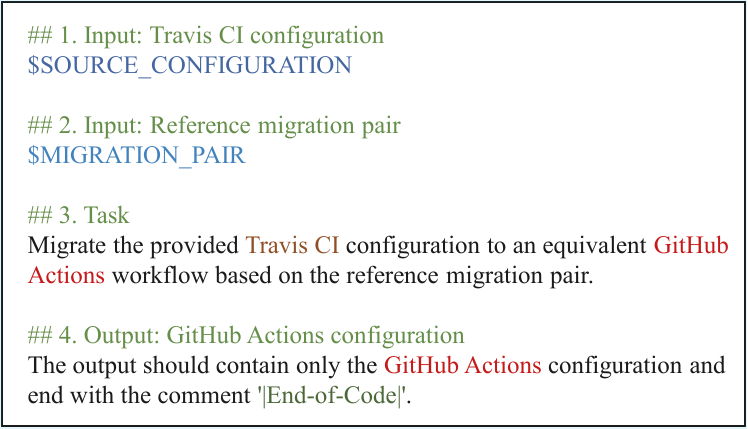}
    \vspace{-17pt}
    \caption{\footnotesize One-shot Prompting} 
    \label{fig:one-shot}
\end{subfigure}
~
\begin{subfigure}[t]{0.33\textwidth}
    \includegraphics[width=\textwidth]{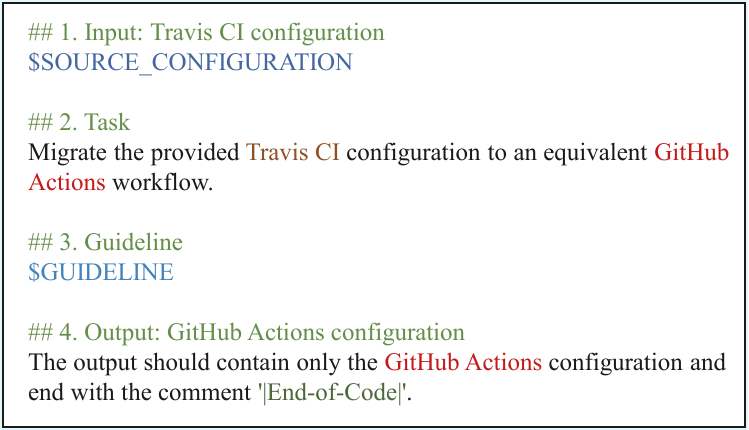}
    \vspace{-17pt}
    \caption{\footnotesize Guideline-Enhanced Prompting}
    \label{fig:guideline-enhanced}
\end{subfigure}
~
\begin{subfigure}[t]{0.33\textwidth}
    \includegraphics[width=\textwidth]{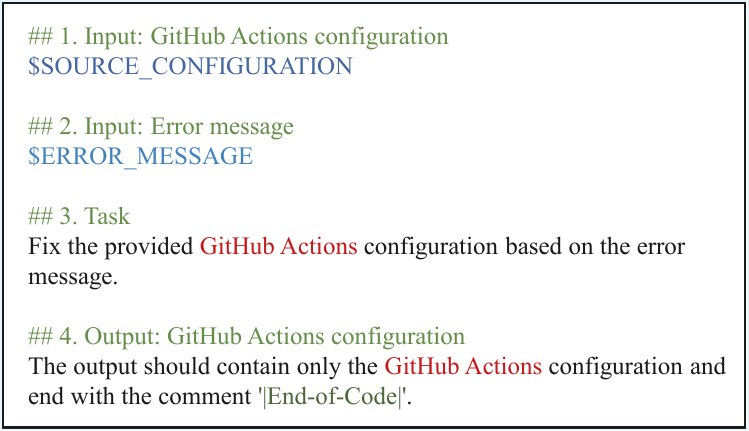}
    \vspace{-17pt}
    \caption{\footnotesize Iterative Refinement } 
    \label{fig:iterative}
\end{subfigure}
\vspace{-10pt}
\caption{Prompt Templates for Three Enhancement Strategies}
\label{fig:enahncement-prompt}
\end{figure}

We adopt three enhancement strategies to investigate their effectiveness in improving CI configuration translation: one-shot prompting, guideline-based prompting, and iterative refinement. The corresponding prompts are illustrated in Figure~\ref{fig:enahncement-prompt}.

\textbf{One-shot Prompting.} In this strategy, the LLM receives a single representative ground-truth translation record. This example illustrates how Travis CI constructs map to their GitHub Actions equivalents, helping the LLM align its output with the expected structure. In practice, for each case, we treated the remaining 228 cases as the candidate ground-truth pool. We then computed textual similarity using term frequency and selected the most similar example as the reference. The selected example, including both the original Travis CI configuration and its corresponding GitHub Actions translation, was inserted into the prompt. 

\textbf{Guideline-based Prompting.} This strategy leverages our taxonomy of translation issues by transforming each issue type into explicit natural language rules to guide the LLM during translation. Unlike one-shot prompting, which relies on providing an example, guideline-based prompting emphasizes generalizable instructions that cover syntax rules, platform-specific differences, environment requirements, and logical constraints. We hypothesize that richer contextual information can improve the quality of generated configurations. For example, to avoid trailing-zero issues (e.g., incorrectly converting \texttt{3.10} into \texttt{3.1}), the guideline explicitly instructs the model to preserve version numbers in string format. The complete guidelines are available on our website \url{https://citranslation.github.io/}.

\textbf{Iterative Refinement.} During workflow execution, the CI platform records the build process and produces detailed outputs, including error messages on failures. This strategy leverages these error messages to iteratively refine the translated configuration. After the LLM generates an initial translation, the workflow is executed on GitHub Actions. If it fails, the error messages are appended to the prompt with the faulty configuration, and the LLM regenerates a corrected version. This cycle continues until one of two termination conditions is met: (1) the workflow for the given case executes successfully, or (2) no additional cases are fixed in the current iteration.

\section{Results and Analyses}\label{sec:results}

\subsection{RQ1: How much effort is required to achieve a successful translation from Travis CI to GitHub Actions?}\label{sec:rq1}

\begin{figure}[htbp]
\centering
\begin{subfigure}[t]{0.33\textwidth}
    \includegraphics[width=\textwidth]{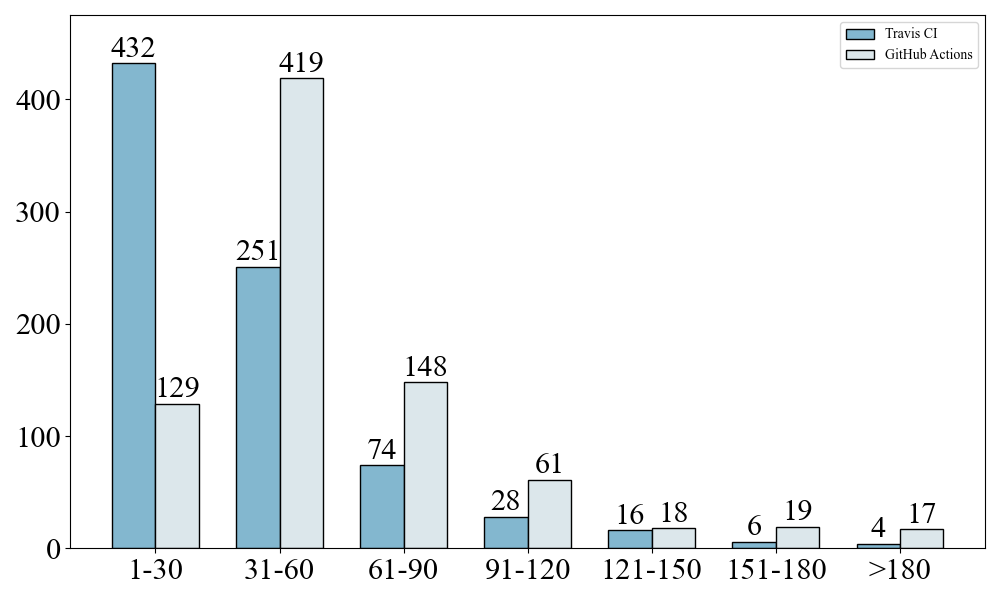}
    \vspace{-17pt}
    \caption{\footnotesize Configuration Size} 
    \label{fig:configuration-size}
\end{subfigure}
~
\begin{subfigure}[t]{0.328\textwidth}
    \includegraphics[width=\textwidth]{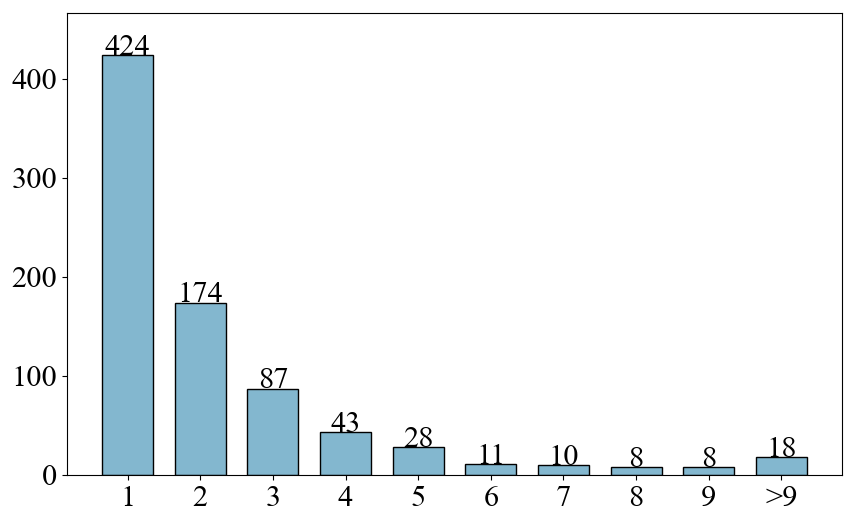}
    \vspace{-17pt}
    \caption{\footnotesize Migration Attempts}
    \label{fig:migration-attempts}
\end{subfigure}
~
\begin{subfigure}[t]{0.33\textwidth}
    \includegraphics[width=\textwidth]{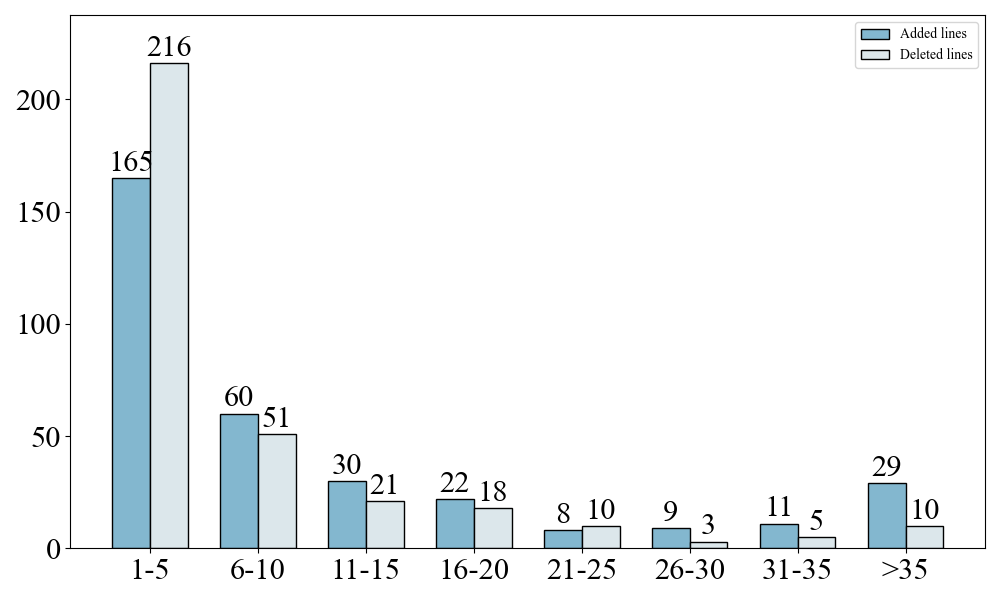}
    \vspace{-17pt}
    \caption{\footnotesize Change Size} 
    \label{fig:change-size}
\end{subfigure}
\vspace{-10pt}
\caption{Effort Analysis of CI Configuration Translation}
\label{fig:effort-analysis}
\end{figure}

Figure~\ref{fig:effort-analysis} summarizes the effort required for successful translations, analyzed from three perspectives: configuration size, migration attempts, and change size. 
Figure~\ref{fig:configuration-size} shows the distribution of configuration sizes across 811 translation records. The $x$-axis denotes the number of configuration lines in Travis CI and GitHub Actions, and the $y$-axis indicates the number of records within each interval (e.g., 1-30 corresponds to the range [1, 30]). Exact counts are annotated above each bar. On average, developers read 38 lines of Travis CI configuration and wrote 58 lines of GitHub Actions configuration. Notably, more than 6.6\% of the translation records required over 120 lines of GitHub Actions configuration.

Figure~\ref{fig:migration-attempts} shows the distribution of commits required for a successful translation. The $x$-axis represents the number of commits, and the $y$-axis indicates the number of records in each interval. Exact counts are annotated above each bar. Among the 811 records, 387 (47.7\%) required multiple commits, and 213 (26.3\%) required three or more, indicating that CI configuration translation is a non-trivial and iterative task. For these multi-commit cases, the migration process lasted an average of \todo{2.5 days} before stabilizing.

Figure~\ref{fig:change-size} focuses on the 387 records requiring multiple commits and illustrates the scale of modifications in follow-up commits after the initial one. The $x$-axis distinguishes between added and deleted lines, and the $y$-axis indicates the number of records in each interval. On average, each follow-up commit involved twelve added and seven deleted lines, while the most complex case included \todo{225 modifications}. We further examined the purpose of these commits: 38\% introduced or removed functional elements, 34\% fixed bugs, and 28\% adjusted style or formatting.

\subsection{RQ2: What types of translation issues occur in LLM-generated configurations?}\label{sec:rq2}

\begin{table}[!t]
\centering
\caption{Issue Types and Frequencies across Different LLMs}
\small
\label{tab:issue-overview}
\begin{tabular}{lccccc}
\toprule
\textbf{Category/Type} & \textbf{GPT-4o} & \textbf{GPT-4o mini} & \textbf{Qwen-3} & \textbf{DeepSeek-Coder} & \textbf{Total}\\
\hline
\rowcolor{gray!20}
\textbf{Syntax Error} & 6 (3\%) & 11 (4\%)& 32 (9\%) & 12 (4\%) & 61 (5\%) \\
\hline
-- Missing Symbol & 0 & 0 & 1 & 4&5 \\
\hline
-- Indentation Error & 3 & 7 & 17 & 3&30 \\
\hline
-- Missing/Misplaced Def. & 0 & 1 & 7 & 1&9 \\
\hline
-- Invalid Value &3&3&7&4&17\\
\hline
\rowcolor{gray!20}
\textbf{Platform Discrepancy} & 97 (43\%) & 72 (28\%) & 56 (16\%) & 131 (45\%) & 356 (32\%) \\
\hline
-- Unsupported Key &23&8 & 4& 31&66 \\
\hline
-- Unsupported Expression &18& 11& 5&18&52 \\
\hline
-- Unsupported Architecture & 3 & 0 & 2 & 1&6 \\
\hline
-- Trailing Zero &10 & 10 & 8& 7&35 \\
\hline
-- Unspecified Default & 5&4&3&35&47\\
\hline
-- \todo{Missing package}&38&39&34&39 & 150\\
\hline
\rowcolor{gray!20}
\textbf{Environment Error} & 64 (28\%) & 59 (23\%) & 76 (22\%) & 85 (29\%) & 284 (25\%) \\
\hline
-- Obsolete Action & 18 & 12& 25 &39 & 94\\
\hline
-- Missing Secret & 46 & 47 & 51 & 46 & 190 \\
\hline
\rowcolor{gray!20}
\textbf{Logic Inconsistency} & 60 (26\%) & 111 (44\%) & 184 (53\%) & 65 (22\%) & 420 (38\%) \\
\hline
-- Trigger Event Misconfig. & 6 & 16 & 4 & 8 & 34\\
\hline
-- Execution Order Error & 7 & 13 & 60 & 14 & 94\\
\hline
-- Condition Misconfig. & 1 & 7 & 3 & 20 & 31 \\
\hline
-- Missing Task & 33&47&94&18 & 192 \\
\hline
-- Redundant Task &13&28&23&5 &69 \\

\bottomrule
\end{tabular}
\end{table}
 

Table~\ref{tab:issue-overview} summarizes the types of issues that occurred during CI configuration translation by LLMs and reports their distributions. The identified issues are grouped into four high-level categories: \textbf{Syntax Error}, \textbf{Platform Discrepancy}, \textbf{Environment Error}, and \textbf{Logic Inconsistency}. Each category is further divided into finer-grained issue types. For each issue type, the table reports the frequency of occurrences across four LLMs. Category rows also report the total number of issues observed within that category, along with their percentage relative to all issues generated by the corresponding model (shown in parentheses). The final column aggregates totals across all models to provide an overall distribution. We describe each issue type in detail below.

\begin{figure}[htbp]
\centering
\begin{subfigure}[t]{0.1838\textwidth}
    \includegraphics[width=1\textwidth]{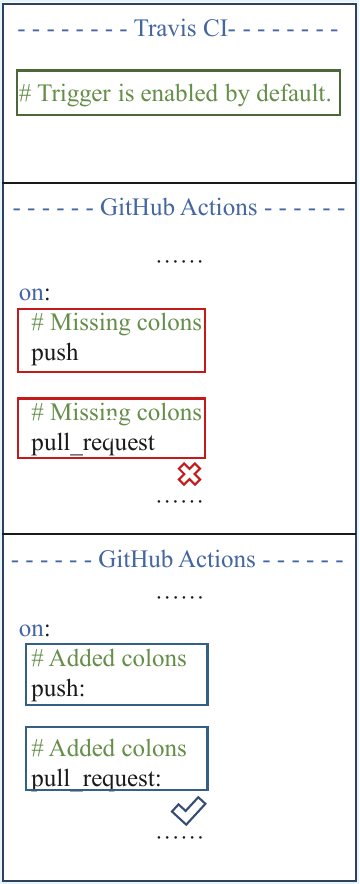}
    \vspace{-17pt}
    \caption{\footnotesize Missing Symbol} 
    \label{fig:missing-symbol}
\end{subfigure}
~
\begin{subfigure}[t]{0.260\textwidth}
    \includegraphics[width=\textwidth]{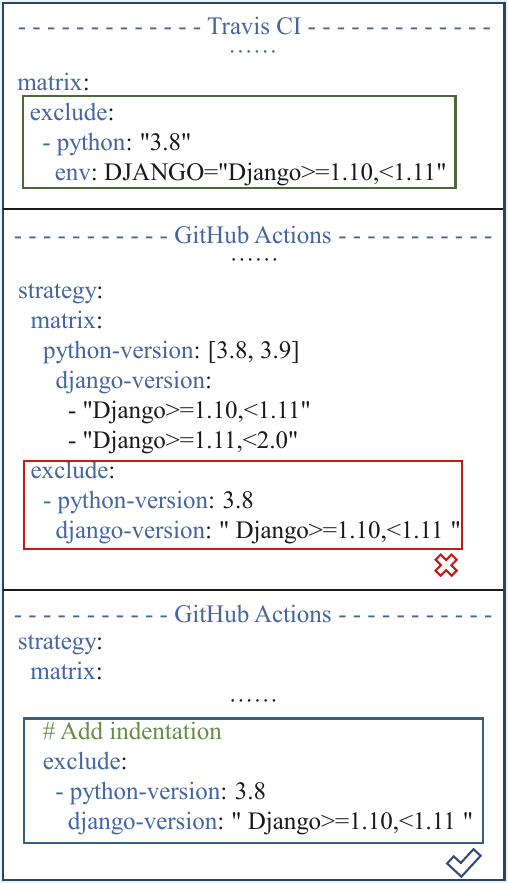}
    \vspace{-17pt}
    \caption{\footnotesize Indentation\_Error}
    \label{fig:indentation-error}
\end{subfigure}
~
\begin{subfigure}[t]{0.2795\textwidth}
    \includegraphics[width=\textwidth]{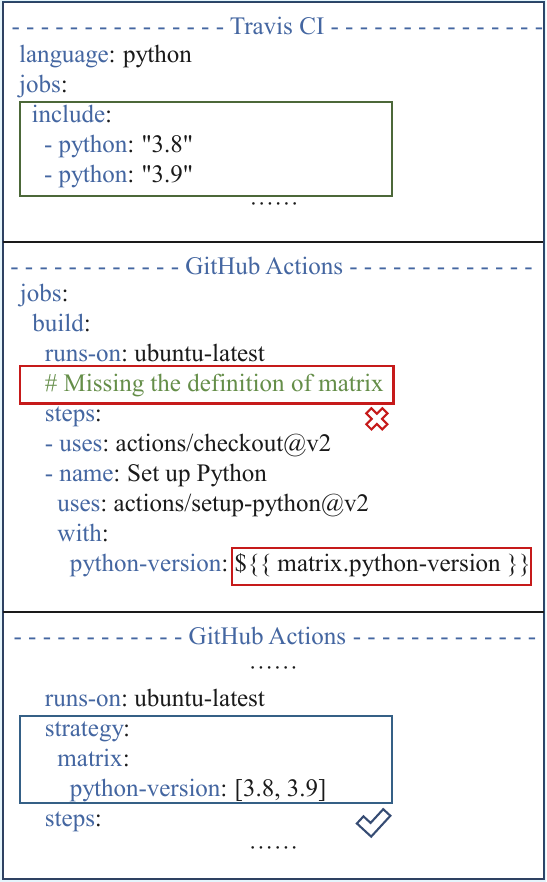}
    \vspace{-17pt}
    \caption{\footnotesize Missing or Misplaced Definition } 
    \label{fig:missing-misplaced-definition}
\end{subfigure}
~
\begin{subfigure}[t]{0.271\textwidth}
    \includegraphics[width=\textwidth]{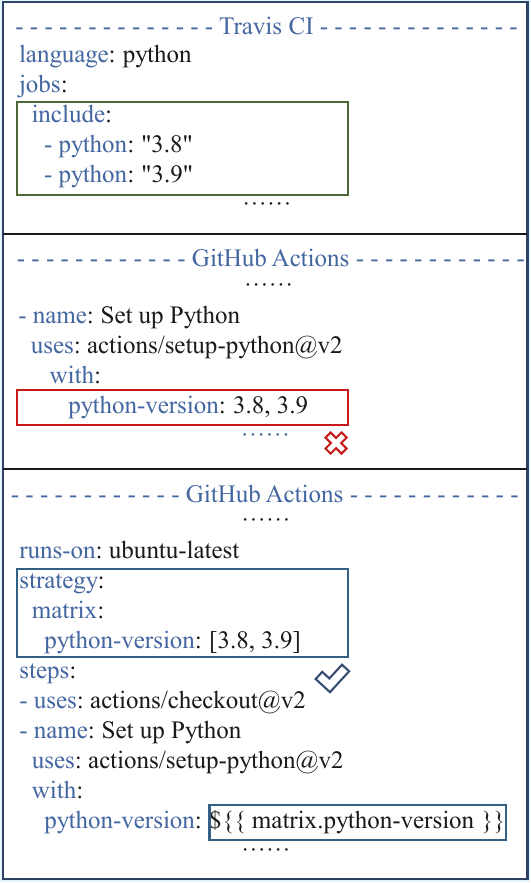}
    \vspace{-17pt}
    \caption{\footnotesize Invalid Value } 
    \label{fig:invalid-value}
\end{subfigure}
\vspace{-10pt}
\caption{Examples of Syntax Errors}
\label{fig:syntax-error}
\end{figure}

\textbf{Syntax Error.} This category comprises four types of issues, all arising from incorrect syntactic or structural configurations. Figure~\ref{fig:syntax-error} presents representative cases for each type.

\begin{itemize}[leftmargin=15pt]

\item \textbf{Missing Symbol.} This error occurs when essential YAML symbols (e.g., colons, hyphens, or quotation marks) are omitted. For example, Figure~\ref{fig:missing-symbol} shows a case where the workflow was intended to be triggered by the events \texttt{push} and \texttt{pull\_request}, but both were written without trailing colons or leading hyphens. Such omissions render the YAML invalid and prevent GitHub Actions from parsing the file. The correct form requires either a colon after each event (e.g., \texttt{push:} and \texttt{pull\_request:}) to declare them as mapping keys, or a leading hyphen (e.g., \texttt{- push}, \texttt{- pull\_request}) to declare them as sequence elements.

\item \textbf{Indentation Error.} Because YAML is whitespace-sensitive, incorrect indentation can cause syntax errors. Figure~\ref{fig:indentation-error} shows an example in which the \texttt{exclude} block is aligned with \texttt{matrix} instead of nested within it. In GitHub Actions, \texttt{exclude} must be defined as a child of \texttt{matrix} to specify which parameter combinations should be omitted. Such misalignment invalidates the configuration. The correct structure requires \texttt{exclude} to be indented under \texttt{matrix}, ensuring that the exclusions are applied correctly.

\item \textbf{Missing or Misplaced Definition.} This error arises when essential configuration definitions are missing or misplaced during translation. Figure~\ref{fig:missing-misplaced-definition} shows an example in which the workflow references the \texttt{matrix.python-version} variable, but the entire \texttt{matrix} definition itself is missing from the translated configuration. Without a declared strategy matrix, GitHub Actions cannot resolve \texttt{matrix.python-version}, rendering the configurations invalid. As a result, the workflow cannot be parsed or executed. The correct configuration requires explicitly defining the \texttt{matrix.python-version} key under \texttt{strategy}, ensuring that all references to \texttt{matrix.python-version} are properly resolved.

\item \textbf{Invalid Value.} This error occurs when the value assigned to a key does not match the expected type or format. Figure~\ref{fig:invalid-value} shows a case in which the key \texttt{python-version} is assigned multiple versions in a comma-separated string. However, the \texttt{                                                                                                   -python} action expects a single version value rather than a list or concatenated values. This mismatch produces an invalid configuration and prevents the workflow from running successfully.

\end{itemize}

\begin{figure}[htbp]
\centering
~
\begin{subfigure}[t]{0.1519\textwidth}
    \includegraphics[width=\textwidth]{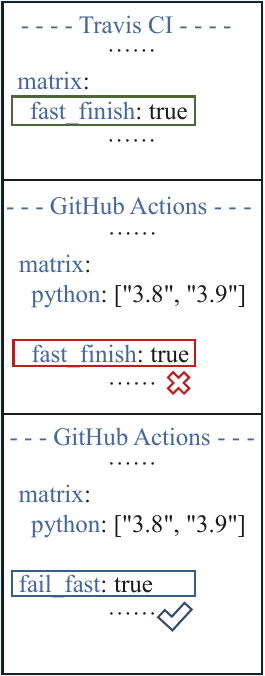}
    \vspace{-17pt}
    \caption{\footnotesize Unsupported Key}
    \label{fig:unsupported-key}
\end{subfigure}
~
~
\begin{subfigure}[t]{0.1559\textwidth}
    \includegraphics[width=\textwidth]{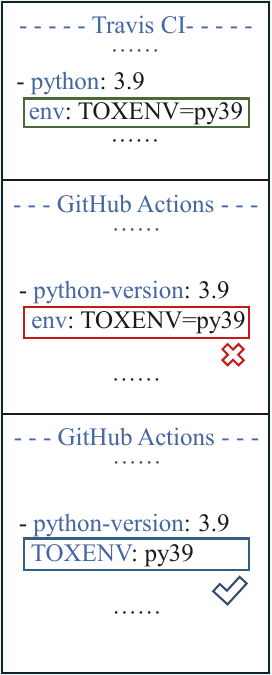}
    \vspace{-17pt}
    \caption{\footnotesize Unsupported Expression}
    \label{fig:unsupported-expression}
\end{subfigure}
~
\begin{subfigure}[t]{0.2355\textwidth}
    \includegraphics[width=\textwidth]{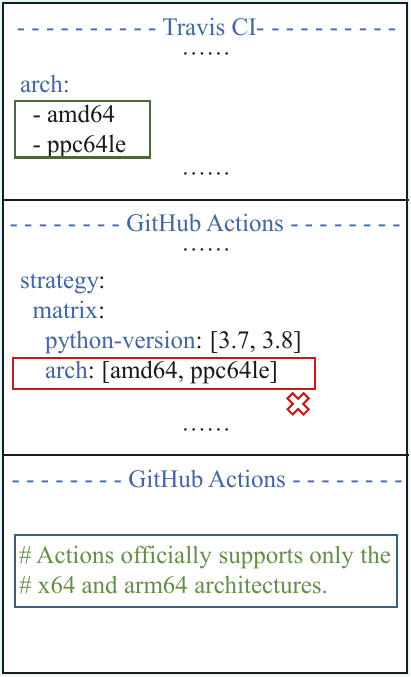}
    \vspace{-17pt}
    \caption{\footnotesize Unsupported Architecture } 
    \label{fig:unsupported-architecture}
\end{subfigure}
~
\begin{subfigure}[t]{0.1905\textwidth}
    \includegraphics[width=\textwidth]{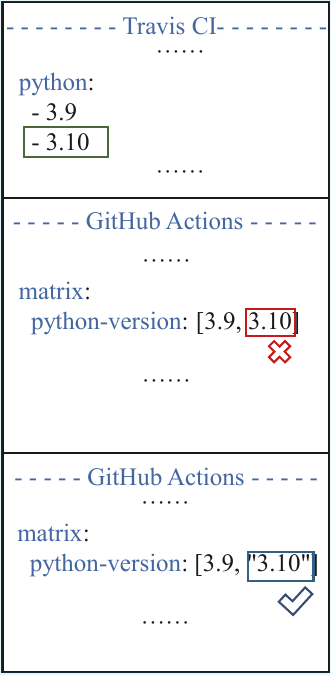}
    \vspace{-17pt}
    \caption{\footnotesize Trailing Zero } 
    \label{fig:trailing-zero}
\end{subfigure}

\begin{subfigure}[t]{0.2615\textwidth}
    \includegraphics[width=\textwidth]{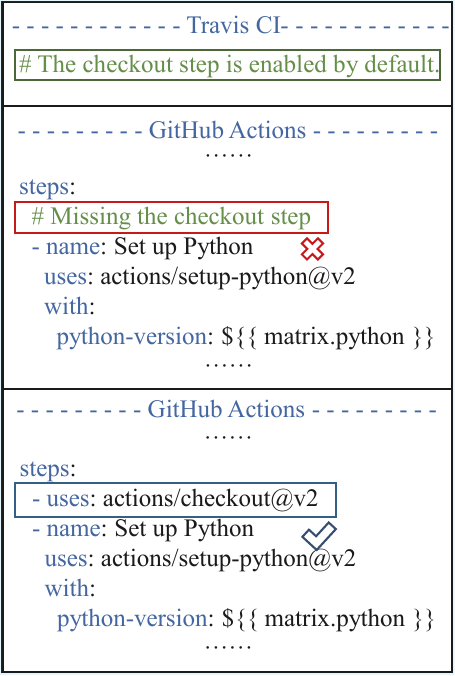}
    \vspace{-17pt}
    \caption{\footnotesize Unspecified Default } 
    \label{fig:unspecified-default}
\end{subfigure}
~
\begin{subfigure}[t]{0.5155\textwidth}
    \includegraphics[width=\textwidth]{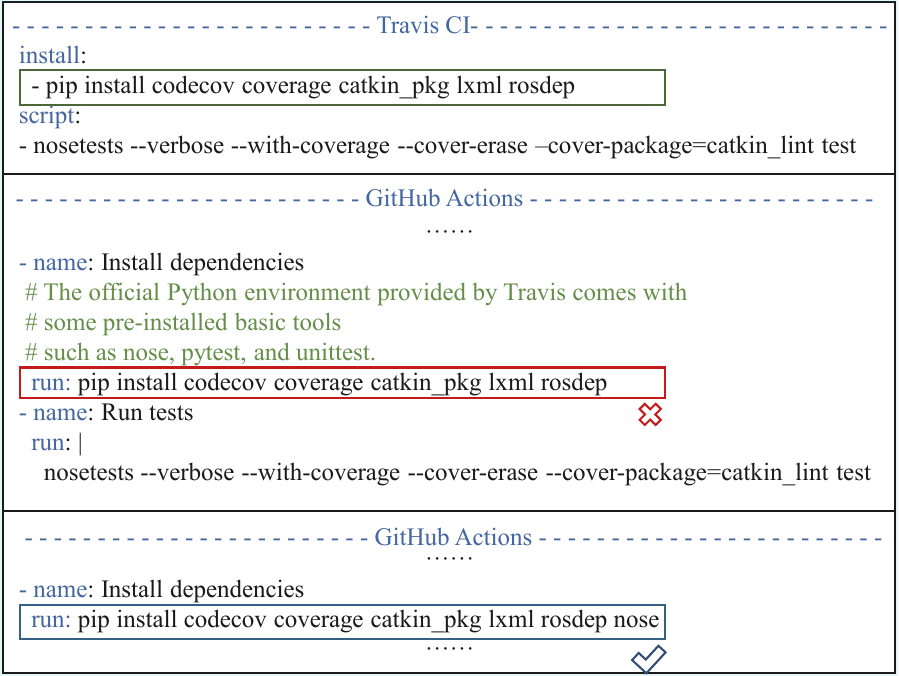}
    \vspace{-17pt}
    \caption{\footnotesize Missing Package}
    \label{fig:missing-package}
\end{subfigure}

\vspace{-10pt}
\caption{Examples of Platform Discrepancy}
\label{fig:platform-discrepancy}
\end{figure}

\textbf{Platform Discrepancy.} This category comprises \todo{six} issue types that stem from differences between Travis CI and GitHub Actions.

\begin{itemize}[leftmargin=15pt]

\item \textbf{Unsupported Key.} This error occurs when Travis CI--specific keys are carried over into GitHub Actions without proper translation. Figure~\ref{fig:unsupported-key} shows an example involving the key \texttt{fast-finish}, which is supported in Travis CI but not recognized in GitHub Actions. The equivalent key in GitHub Actions is \texttt{fail-fast}, which controls whether the remaining jobs in a matrix should be canceled once one of them fails. 

\item \textbf{Unsupported Expression.} This error occurs when Travis CI expressions are carried over into GitHub Actions without proper adaptation, resulting in invalid configurations. As shown in Figure~\ref{fig:unsupported-expression}, one example arises in environment variable declarations: in Travis CI, environment variables can be expressed as assignment strings (e.g., \texttt{env: TOX\_ENV=flake8}), whereas GitHub Actions requires key–value pairs (e.g., \texttt{TOX\_ENV: flake8}). If the Travis format is used directly, GitHub Actions interprets the entire string literally (e.g., \texttt{TOX\_ENV=flake8}), preventing the intended environment variable from being correctly recognized.

\item \textbf{Unsupported Architecture.} This error occurs when configurations specify hardware architectures that are valid in Travis CI but unsupported in GitHub Actions. Figure~\ref{fig:unsupported-architecture} illustrates a case in which the build matrix includes the value \texttt{ppc64le} under the \texttt{arch} parameter. While Travis CI supports a wider range of architectures, GitHub Actions currently allows only \texttt{x64} and \texttt{arm64}. As a result, specifying any other architecture produces an invalid configuration and prevents the workflow from executing.

\item \textbf{Trailing Zero.} This error occurs when \todo{version numbers} containing trailing zeros are incorrectly parsed or represented during translation. Figure~\ref{fig:trailing-zero} shows an example where the intended Python versions are \texttt{[3.10, 3.11]}. However, the value \texttt{3.10} is misinterpreted as \texttt{3.1}, which changes the actual runtime environment. To avoid this issue, version numbers with trailing zeros must be enclosed in quotes (e.g., "3.10") to correct parsing. Notably, this problem does not occur in Travis CI.

\item \textbf{Unspecified Default.} This error occurs when settings that are implicit in one platform must be explicitly configured in the other. For example, as shown in Figure~\ref{fig:unspecified-default}, Travis CI automatically checks out the repository code without requiring explicit instructions. In contrast, GitHub Actions requires this step to be declared explicitly by invoking \texttt{actions/checkout@v2}.

\item \textbf{Missing Package.} This error occurs when packages implicitly available on one platform are not explicitly installed on the other. As shown in Figure~\ref{fig:missing-package}, the \texttt{nose} package is pre-installed in the Travis CI by default but must be explicitly installed in GitHub Actions.


\end{itemize}

\begin{figure}[!t]
\centering
~
\begin{subfigure}[t]{0.4\textwidth}
    \includegraphics[width=\textwidth]{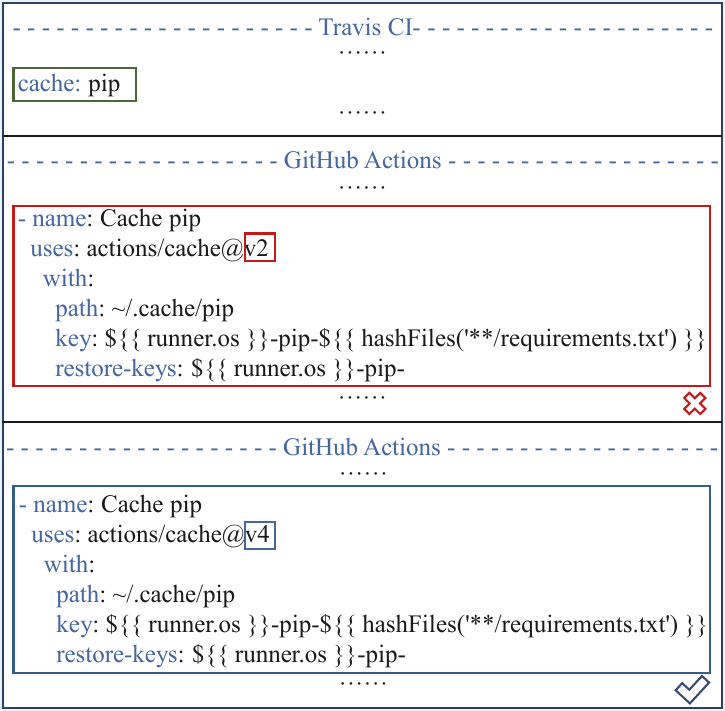}
    \vspace{-17pt}
    \caption{\footnotesize Obsolete Action}
    \label{fig:obsolete-action}
\end{subfigure}
~
~
\begin{subfigure}[t]{0.339\textwidth}
    \includegraphics[width=\textwidth]{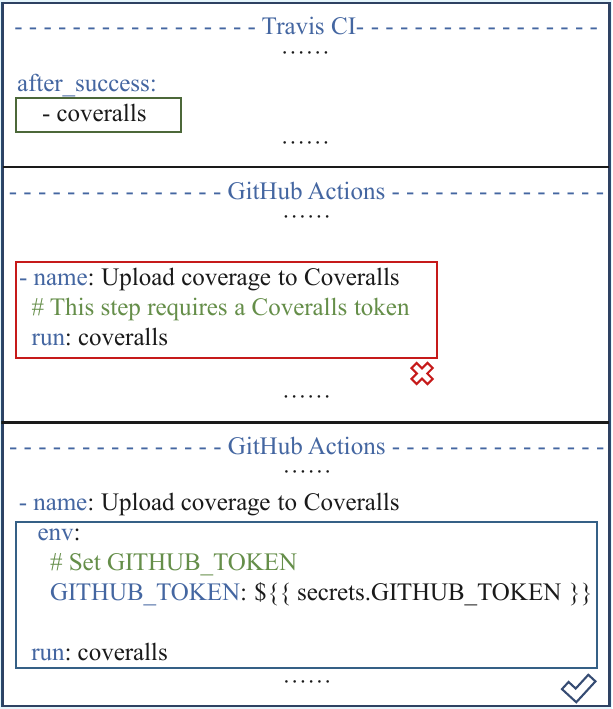}
    \vspace{-17pt}
    \caption{\footnotesize Missing Secret}
    \label{fig:missing-secret}
\end{subfigure}

\vspace{-10pt}
\caption{Examples of Environment Error}
\label{fig:environment-error}
\end{figure}

\textbf{Environment Error.} This category involves two types of issues caused by improperly configured environments, such as referencing obsolete actions or failing to provide required credentials.

\begin{itemize}[leftmargin=15pt]

\item \textbf{Obsolete Action.} This error occurs when the translated workflow references actions that have become obsolete or deprecated. As shown in Figure~\ref{fig:obsolete-action}, the workflow uses \texttt{actions/cache@v2}, which is no longer supported since only \texttt{v3} and later versions are maintained. The correct approach is to update references to the latest supported version (e.g., \texttt{actions/cache@v4}).

\item \textbf{Missing Secret.} This error occurs when required credentials or tokens are not properly configured in GitHub Actions. For example, Figure~\ref{fig:missing-secret} shows a workflow step intended to upload coverage reports to Coveralls. In Travis CI, the token might have been implicitly available or injected via environment variables, but in GitHub Actions it must be explicitly provided through the secrets context (e.g., \texttt{secrets.GITHUB\_TOKEN}). If the secret is missing, the Coveralls command will be unable to authenticate, causing the upload step to fail.

\end{itemize}

\begin{figure}[htbp]
\centering
~
\begin{subfigure}[t]{0.1420\textwidth}
    \includegraphics[width=\textwidth]{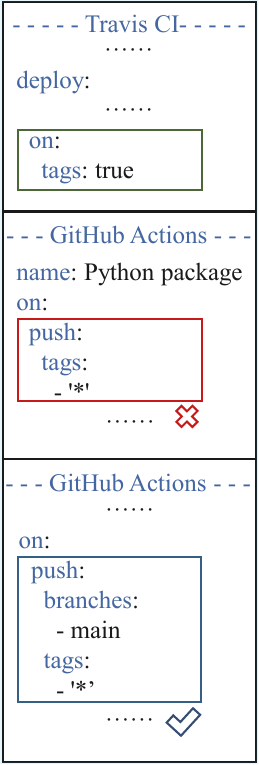}
    \vspace{-17pt}
    \caption{\footnotesize Trigger Event Misconfiguration}
    \label{fig:trigger-event-misconfiguration}
\end{subfigure}
~
\begin{subfigure}[t]{0.1599\textwidth}
    \includegraphics[width=\textwidth]{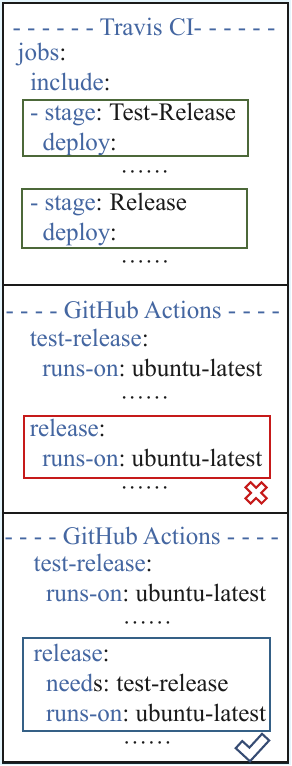}
    \vspace{-17pt}
    \caption{\footnotesize Execution Order Error}
    \label{fig:execution-order-error}
\end{subfigure}
~
\begin{subfigure}[t]{0.2279\textwidth}
    \includegraphics[width=\textwidth]{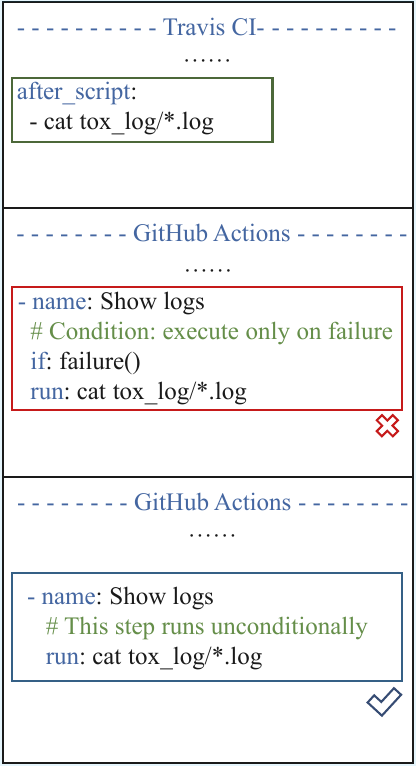}
    \vspace{-17pt}
    \caption{\footnotesize Condition Misconfiguration}
    \label{fig:condition-misconfiguration}
\end{subfigure}
~
\begin{subfigure}[t]{0.2423\textwidth}
    \includegraphics[width=\textwidth]{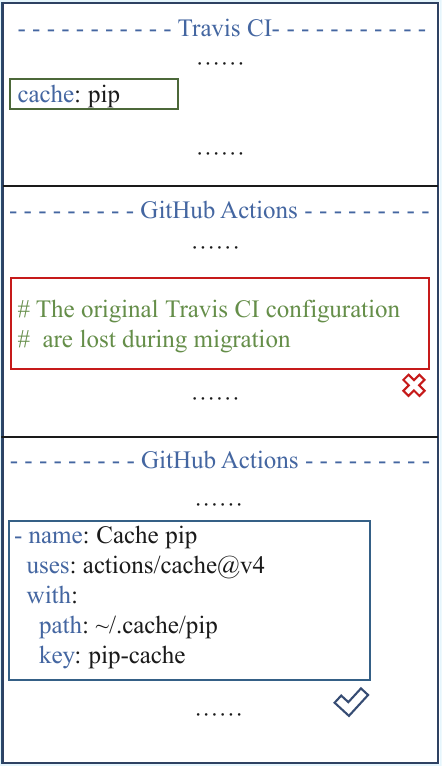}
    \vspace{-17pt}
    \caption{\footnotesize Missing Task}
    \label{fig:missing-task}
\end{subfigure}
~
\begin{subfigure}[t]{0.2282\textwidth}
    \includegraphics[width=\textwidth]{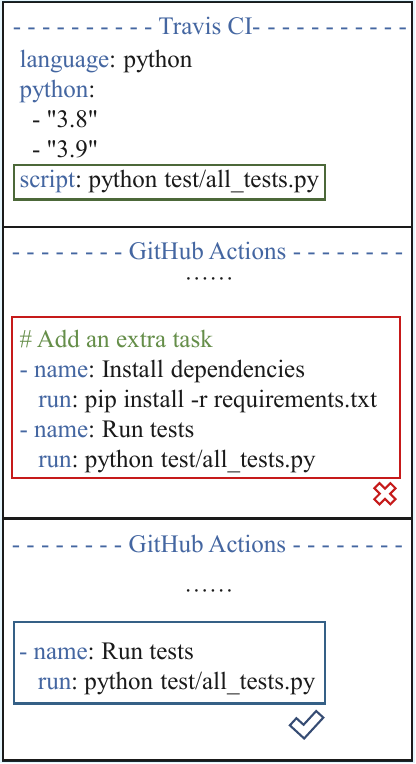}
    \vspace{-17pt}
    \caption{\footnotesize Redundant Task}
    \label{fig:redundant-task}
\end{subfigure}

\vspace{-10pt}
\caption{Examples of Logic Inconsistency}
\label{fig:logic-nconsistency}
\end{figure}

\textbf{Logic Inconsistency.} This category covers \todo{five} types and involves issues where the logical semantics of the original Travis CI configurations are not preserved in the translated GitHub Actions workflow.

\begin{itemize}[leftmargin=15pt]

\item \textbf{Trigger Event Misconfiguration.} This issue occurs when trigger events are incorrectly translated, causing workflows to execute under unintended conditions. A representative case is shown in Figure~\ref{fig:trigger-event-misconfiguration}. In the original Travis CI configuration, the deploy stage is triggered only when a tagged commit is pushed, whereas the build stage runs on every push. After translation, the GitHub Actions workflow was configured with \texttt{push} restricted to \texttt{tags: '*'}, meaning that normal (non-tag) pushes no longer trigger the build stage and only tagged pushes trigger both build and deploy. As a result, the semantics of the CI workflow are altered.


\item \textbf{Execution Order Error.} This issue occurs when the intended execution order of jobs or steps is not preserved during translation, leading to incorrect sequencing or unintended concurrency in GitHub Actions. As shown in Figure~\ref{fig:execution-order-error}, the original Travis CI configuration defined a clear sequential order: first run \texttt{test-release}, and then \texttt{release}. However, after translation into GitHub Actions, the dependencies between jobs were not explicitly specified using the \texttt{needs} keyword. Without these dependency constraints, GitHub Actions defaults to running jobs concurrently, which breaks the intended execution order and can cause premature deployment or inconsistent results. The correct translation should explicitly declare dependencies (e.g., \texttt{release} needs \texttt{test-release}) to preserve the sequential execution semantics of the original Travis CI configuration.

\item \textbf{Condition Misconfiguration.} This category refers to issues caused by incorrectly introduced or misplaced \texttt{if} conditions during translation. As shown in Figure~\ref{fig:condition-misconfiguration}, additional \texttt{if} conditions may be inadvertently introduced. In the original configuration, log outputs were always displayed regardless of success or failure. After translation, however, an extra condition \texttt{if: failure()} was inserted, restricting log output to failure cases only.

\item \textbf{Missing Task.} This issue occurs when tasks defined in the original Travis CI configuration are omitted during translation, leading to missing functionality in the generated GitHub Actions workflow. Figure~\ref{fig:missing-task} shows an example in which the caching step (\texttt{cache: pip}) in Travis CI was lost during migration and should have been explicitly reproduced in GitHub Actions using \texttt{actions/cache}.

\item \textbf{Redundant Task.} This issue occurs when tasks absent from the original Travis CI configuration are incorrectly introduced into the GitHub Actions workflow during translation. As shown in Figure~\ref{fig:redundant-task}, an extra step of running \texttt{pip install -r requirements.txt} was added. Because this step was not present in the original configuration, its inclusion constitutes a redundancy introduced during translation.

\end{itemize}

Among these categories, logic inconsistency is the most prevalent, accounting for more than one-third of all issues \todo{(38\%)} and ranking as the dominant issue category for GPT-4o mini and Qwen-3, and third for GPT-4o and DeepSeek-Coder. Typical examples include missing tasks (192 cases), execution order errors (94 cases), and redundant tasks (69 cases). These findings indicate that LLMs often struggle to preserve the intended execution semantics during translation, resulting in workflows that omit necessary tasks, introduce redundant tasks, or execute tasks in the wrong order, thereby deviating from the original design.

The second most frequent category is platform discrepancy issues, which account for 32\% of all errors, primarily arising from missing packages (150 cases), unsupported keys (66 cases), and unsupported expressions (52 cases). These discrepancies rank first for GPT-4o and DeepSeek-Coder, second for GPT-4o mini, and third for Qwen-3. This highlights the inherent differences between the CI domain-specific languages of Travis CI and GitHub Actions. Although both platforms use YAML for configuration, their syntax and semantics differ substantially. Many steps implicit in one platform must be explicitly defined in the other. In addition, numerous configuration features supported by one platform require adaptation during migration to the other. These platform-specific differences create substantial translation challenges because LLMs often struggle to identify and map features across platforms, leading to missing or incorrect mappings.

Environment error (25\%), with the majority of errors caused by missing secrets (190 cases). This reveals a critical challenge related to credentials: missing secrets block access to external resources such as APIs or third-party services, which are essential for CI workflows. The frequent occurrence of missing secret issues suggests that LLMs may not fully account for platform-specific authentication mechanisms.

Syntax errors account for only 5\% of the total issues, including indentation errors (30 cases), invalid values (17 cases), and missing/misplaced definitions (9 cases). In terms of distribution across models, syntax errors consistently rank fourth for all four LLMs, indicating that they are the least common type of issue. This suggests that LLMs are generally competent at producing syntactically valid YAML because the language’s surface structure is relatively simple and strongly reinforced during pretraining. However, the presence of these errors still exposes weaknesses. Indentation mistakes are particularly critical because YAML is indentation-sensitive, and even a minor misalignment can invalidate the entire configuration.

Across the evaluated models, GPT-4o exhibited the fewest issues (227 cases), followed by GPT-4o mini (253), DeepSeek-Coder (293), and Qwen-3 (348). A similar trend appears for BSR: GPT-4o achieved the highest score (25.8\%), followed by GPT-4o mini (24.5\%), Qwen-3 (22.7\%), and DeepSeek-Coder (16.2\%). These results indicate that larger and more advanced models handle CI configuration translation more effectively, producing fewer issues and more executable workflows.

\subsection{RQ3: To what extent can enhancement strategies improve the performance of LLM-based CI configuration translation?}\label{sec:rq3}

According to the RQ2 results with the basic prompt, the baseline performance of LLMs in CI configuration translation is unsatisfactory and prone to multiple issues. Even the best-performing model, GPT-4o, achieved a Build Success Rate (BSR) of only 25.8\% (59 successful translations). We therefore focus on GPT-4o and evaluate three enhancement strategies to assess whether translation quality can be further improved. For comparison, we also include the official migration tool, \textit{Importer}, as a baseline. We did not include \textit{CIMig}~\cite{rzig2024example}, as it contains rules specifically designed for Java projects and is therefore not applicable to the Python projects in our dataset. Table~\ref{tab:performance-overview} summarizes the results. The table reports results for the Importer baseline, GPT-4o with a Basic prompt as the LLM baseline, three enhancement strategies (One-shot, Guideline, and IR for Iterative Refinement), and their combination (Guideline+IR).

In terms of the BSR, the Importer tool achieved 18.3\% (42 successful translations), which is lower than all GPT-4o–based strategies. This highlights the performance gap between rule-based and LLM-based approaches. Compared with the LLM baseline, one-shot prompting reduced performance to 21.0\% (48 successful translations), indicating that providing a single example did not improve accuracy and even hindered generalization. Guideline-based prompting substantially improved performance, raising the BSR to 40.2\% (92 successful translations), a 55.9\% relative improvement over the LLM baseline. Iterative refinement achieved the best performance, with a BSR of 68.6\% (157 successful translations), representing a 2.7 times improvement over the LLM baseline.

Because both guideline-based prompting and iterative refinement yielded substantial improvements over the LLM baseline, but each targeted different aspects of the translation challenge, we explored their combination. Guideline-based prompting reduces issues by providing explicit rules on syntax, platform conventions, environment requirements, and logic constraints, whereas iterative refinement corrects mistakes by leveraging error messages as feedback across multiple iterations. Motivated by their complementary strengths, we combined them into the Guideline+Iterative Refinement (Guideline+IR) strategy. In practice, GPT-4o was first guided by guideline-enhanced prompting during the initial translation, and subsequent iterations used iterative refinement to progressively refine the configuration until either the workflow executed successfully or no additional cases were fixed in the current round. As a result, this combined strategy produced the best overall result, achieving a BSR of 75.5\% (173 successful translations), nearly three times higher than the LLM baseline and more than four times higher than the rule-based Importer. These findings demonstrate that structured guidance and iterative feedback are not only effective individually but also highly synergistic when combined.

\begin{figure}[htbp]
\centering
\begin{subfigure}[t]{0.5\textwidth}
    \includegraphics[width=\textwidth]{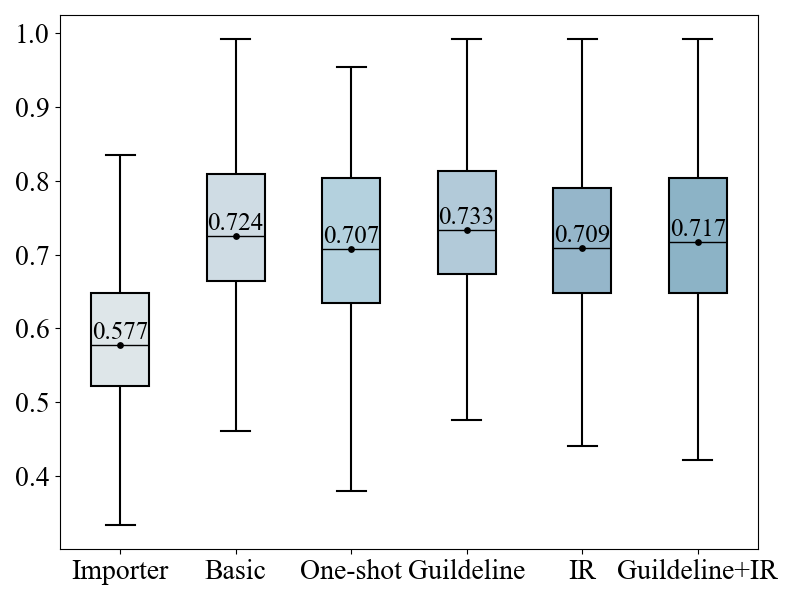}
    \vspace{-17pt}
    \caption{\footnotesize Cosine Similarity} 
    \label{fig:Cosine Similarity}
\end{subfigure}
~
\begin{subfigure}[t]{0.5\textwidth}
    \includegraphics[width=\textwidth]{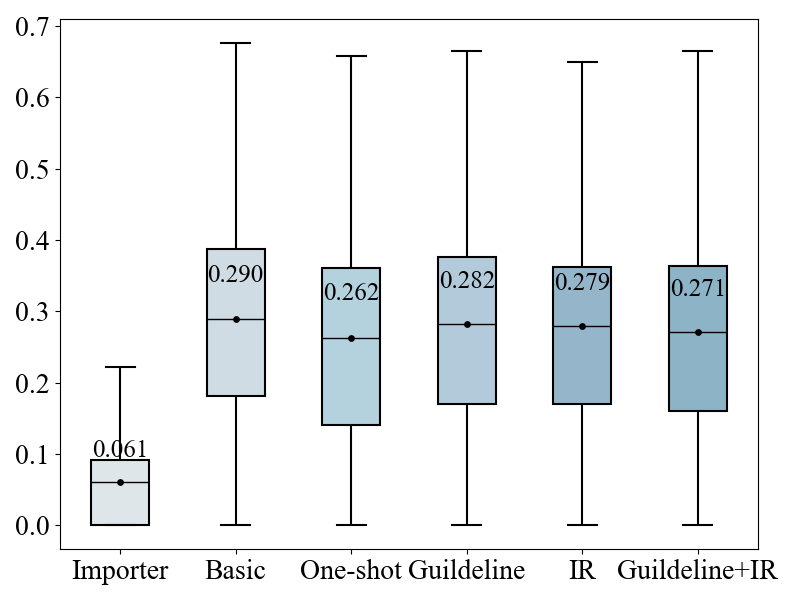}
    \vspace{-17pt}
    \caption{\footnotesize CrystalBLEU}
    \label{fig:CrystalBLEU score}
\end{subfigure}

\vspace{-10pt}
\caption{Results of Similarity-Based Metrics}
\label{fig:similarity-metrics}
\vspace{-19pt}
\end{figure}

For similarity-based metrics, we report the median values of Cosine Similarity (CS) and CrystalBLEU (CB), and present their distributions as box plots in Figure~\ref{fig:similarity-metrics}. For CS, the highest score was achieved by guideline-based prompting, reaching 73.3\%. For CB, the highest score was obtained with the baseline prompt, at 29.0\%. Surprisingly, these metrics do not align with BSR. As noted in prior work~\cite{pan2024lost}, similarity-based metrics may assign high scores to outputs that are textually close to the reference yet still fail at runtime. Therefore, although CS and CB provide complementary insights into surface-level similarity, execution-based evaluation such as BSR should be regarded as the primary measure of translation quality because it more faithfully reflects practical usability in real CI environments.


 \vspace{-5pt}
\begin{table}[ht]
\small
\centering
\caption{Performance of Importer and GPT-4o under Different Prompting Strategies}
\vspace{-10pt}
\label{tab:performance-overview}
\begin{tabular}{lcccccc}
\toprule
\textbf{Metric} & \textbf{Importer} & \textbf{Basic} & \textbf{One-shot} & \textbf{Guideline} & \textbf{IR} & \textbf{Guideline+IR}\\
\hline
\textbf{BSR} & 18.3 (42) & 25.8 (59) & 21.0 (48) & 40.2 (92) & 68.6 (157) & \textbf{75.5 (173)} \\
\hline
\textbf{CS} & 57.7 & 72.4 & 70.7 & \textbf{73.3} & 72.5 & 71.7\\
\hline
\textbf{CB} & 6.1 & \textbf{29.0} & 26.2 & 28.2 & 27.9 & 27.1\\
\bottomrule
\end{tabular}
\vspace{-15pt}
\end{table}



\section{Threats to Validity}\label{sec:threats}

Our work is subject to several potential threats to validity, which we summarize in three aspects.  

\textbf{Internal Validity.} First, to evaluate LLM-based migration, each configuration was translated and executed only once, with the temperature set to 0 whenever possible. Although this setup improves reproducibility, it may underestimate the potential variability of model outputs. Second, manual annotation of translation issues inevitably involved human judgment. To mitigate subjectivity, we applied an open coding procedure, conducted independent labeling by two experienced authors. Disagreements were resolved through consultation with a third author. 


\textbf{External Validity.} First, our study focuses exclusively on migrations from Travis CI to GitHub Actions, motivated by the practical prevalence of this migration pattern. We cannot guarantee that our results generalize to other CI platforms or migration scenarios. However, we believe that some of the findings are broadly applicable.
Second, our dataset contains only Python projects because Python is among the most widely used programming languages~\cite{tiobe-com}. The challenges we identify are not specific to Python and should generalize to other programming languages. Third, our experiments are based on two datasets containing 811 and 229 migration records. Although larger or more diverse datasets may reveal additional patterns, we believe these datasets provide a representative view of CI configuration translation. Moreover, because GitHub Actions supports workflows that can be split across multiple files, we focused on one-to-one translations to simplify the analysis, leaving one-to-many scenarios for future work. Finally, given the large number of available LLMs, we evaluated four representative models (i.e., GPT-4o, GPT-4o mini, Qwen-3, and DeepSeek-Coder), covering both general and code-specific types as well as proprietary and open-source settings.

\textbf{Construct Validity.} The main threat arises from the properties of the evaluation metrics. For RQ1, we used configuration size, migration attempts, and change size to approximate migration effort. Together, these metrics capture both the overall complexity of the final configuration and the incremental cost during the migration process. Similarly, to evaluate translation performance, we selected three metrics that reflect static aspects (Cosine Similarity and CrystalBLEU) and dynamic aspects (Build Success Rate).


\section{Related Work}\label{sec-related}

\textbf{Configuration in CI.} Hilton et al.~\cite{hilton2017trade} interviewed 16 developers to investigate the barriers and needs of using CI. They found that configuring CI services is a common and significant challenge, as it is time-consuming and requires a steep learning curve. Even simple workflows demand considerable configuration effort. 
Pinto et al.~\cite{pinto2018work} conducted a qualitative survey with 158 CI users to explore their understanding of fundamental CI concepts, reasons for build breakage, and the benefits and problems of CI. Their findings corroborated Hilton et al.~\cite{hilton2017trade}, with 31 participants reporting difficulties in configuring CI build environments.
Zampetti et al.~\cite{zampetti2023continuous} further investigated the challenges and barriers faced in developing Cyber-Physical Systems. Their study revealed that developers often fail to properly configure CI services due to insufficient knowledge.
In addition, several studies have focused specifically on CI configuration issues~\cite{gallaba2018use, vassallo2020configuration, zhang2022buildsonic, khatami2024catching}. These studies show that CI services are often suboptimally configured, negatively impacting their correctness, maintainability, performance, and security. These studies show that configuring CI is a non-trivial task, motivating us to focus on the automated solutions for CI configuration translation to reduce the developers' burden.


\textbf{Code Translation.} Code translation is the process of transforming code from one programming language to another. Driven by the growing complexity of software systems and the increasing demand for multi-language development, code translation plays a crucial role in reducing maintenance effort and improving non-functional properties (e.g., reliability, security, and performance) of software systems~\cite{chen2025systematic, pan2024lost}. To support this process, several automated techniques have been proposed, ranging from program analysis approaches~\cite{ling2022rust} to learning-based methods~\cite{nguyen2014migrating}. In particular, the emergence of LLMs demonstrates significant potential in software engineering tasks~\cite{zheng2025towards}. Recent studies have evaluated the effectiveness of LLMs in code translation. Pan et al.~\cite{pan2024lost} conducted a large-scale empirical study using 1,700 code samples across five programming languages: C, C++, Go, Java, and Python. Based on these samples, they examined the ability of LLMs in code translation and identified 15 categories of translation bugs, including syntactic and semantic differences, missing dependencies, incorrect logic implementations, model-specific limitations, and data-related bugs. To mitigate these problems, they further proposed a prompt-crafting approach based on the symptoms of erroneous translations, which improves performance by 5.5\% on average. Yang et al.~\cite{yang2024exploring} focused on five recent LLMs and investigated their issues in code translation. They found that most translation failures are caused by a lack of comprehension of source programs, missing clear instructions on I/O types in translation, and ignoring discrepancies between source and target programs. To address these issues, they proposed \textit{UniTrans}, a framework that enhances code translation by automatically generating a series of test cases to guide and validate the translation process. While these studies examined cross-language code translation, CI configuration presents distinct characteristics.

\textbf{CI Migration.} 
Golzadeh et al.~\cite{golzadeh2022rise} conducted a nine-year longitudinal empirical study to investigate the rapidly evolving CI landscape. They focused on seven popular CI services to examine patterns of co-usage and migration. One in four repositories had used at least two different CI services, and they observed 14,219 migrations across 13,083 repositories. Rostami Mazrae et al.~\cite{rostami2023usage} conducted an in-depth qualitative study by interviewing 22 experienced software practitioners about their usage, co-usage, and migration across 31 CI/CD tools. They identified the main reasons for co-using multiple CI/CD tools and for migrating from one tool to another. These studies highlight the prevalence of CI migration. Rzig et al.~\cite{rzig2024example} proposed an example-based mining approach that extracts translation rules and configuration patterns from existing migration examples and applies them to reproduce similar migrations in new contexts. Unlike these studies, our work investigates the potential of LLMs to automatically translate CI configurations. While Nazmul Hossain et al.~\cite{nazmul2025cigrate} outlined a research plan for an LLM-based framework to automatically migrate CI configurations, they only presented the proposal without reporting any results. Moreover, their plan does not address key aspects such as the effort involved in CI migration, the types of translation issues that may arise, or the role of guidelines and iterative refinement in improving translation performance.

\section{Conclusion}\label{sec:conclusion}

CI migration is a common and non-trivial process. In this paper, we investigate the effort required for CI migration, the types of issues that occur in LLM-based translation, and their distribution. We further evaluate three enhancement strategies to improve translation performance. Our results show that combining guideline-based prompting with iterative refinement significantly enhances translation quality. We plan to explore more advanced LLM techniques and extend the scenario from one-to-one to one-to-many translation.

\section{Data Availability}
The datasets and results of this paper are publicly available at \url{https://citranslation.github.io/}.

\bibliographystyle{ACM-Reference-Format}
\bibliography{references}

\end{document}